\input harvmac
\lref\Cory{D.G. Cory, A.F. Fahmy, T.H. Havel, in PhysComp96,
(ed.) T.Toffoli, M.Biafore and J.Leao (New England Complex Systems 
Inst.,1996),
Proc. Nat. Acad. Sci. 94, 1634 (1997).}
\lref\GershA{N.A. Gershenfeld, I.L. Chuang,
Science 275, 350 (1997).}
\lref\GershB{N.A. Gershenfeld, I.L. Chuang, S. Lloyd,
ibid., 134 (1996).}
\lref\GershC{N.A. Gershenfeld, I.L. Chuang, M. Kubinec,
Phys. Rev. Lett. 80, 3408 (1998).}
\lref\Chuan{I.L Chuang, L.M.K. Vandersypen, X. Zhou, D.W. Leung, S. Lloyd,
Nature 393, 143 (1998).}
\lref\JonesA{J.A. Jones, M. Mosca,
J. Chem. Phys. 109, 1648 (1998).} 
\lref\JonesB{J.A. Jones, M. Mosca, R.H. Hansen,
Nature 393, 344 (1998).} 
\lref\mooij{J.E. Mooij, T.P. Orlando, L. Levitov, L. Tian, C.H. van der 
Wal,
S. Lloyd, Science 285, 1036 (1999).}
\lref\orlando{T.P. Orlando, J.E. Mooij, L. Tian, C.H. van der Wal,
L.S. Levitov, S. Lloyd, J.J. Mazo, Phys. Rev. B60, 15398 (1999).}
\lref\ioffe{L.B. Ioffe, V.B. Geshkenbein, M.V. Feigelman, A.L. Fauchere,
G. Blatter, Nature 398, 679 (1999).}
\lref\girvin{S.M. Girvin, Les Houches Lectures 1998, cond-mat/9907002.}
\lref\susskind{L. Susskind, "The quantum hall fluid and non-commutative
chern simons theory", hep-th/0101029.}
\lref\dunne{G.V. Dunne, Les Houches Lectures 1998, hep-th/9902115.} 
\lref\moon{K. Moon, H. Mori, K. Yang, L. Belkhir, S.M. Girvin,
A.H. MacDonald, L. Zheng, D. Yoshioka, Phys. Rev. B54, 11644 (1996).}
\lref\barenco{A. Barenco, C.H. Bennett, R. Cleve, D.P. DiVincenzo,
N. Margolus, P. Shor, T. Sleator, J. Smolin, H. Weinfurter,
Phys. Rev. A52, 3457 (1995).}
\lref\divi{D.P. DiVincenzo, D. Bacon, G. Burkard, J. Kempe, K.B. Whaley,
quant-ph/0005116.}
\lref\wen{X.G.Wen, A.Zee, Phys. Rev. Lett 69, 1811 (1992).
Phys. Rev. B47 (1992).}
\lref\ezawa{Z.F.Ezawa, A.Iwazaki, Phys. Rev. B48 15189 (1993).} 
\lref\sondhi{S.L.Sondhi, A.Karlhede, S.A. Kivelson, E.H. Rezayi,
Phys. Rev. B47, 16419 (1993).}
\lref\privman{V. Privman, I.D. Vagner, G. Kventsel, Phys. Lett. A239 
(1998).}
\lref\spiA {I.B. Spielman et al. Phys Rev. Lett. 84, 5808 (2000).}
\lref\spiB {I.B. Spielman, J.P. Eisenstein, L.N. Pefeiffer, K.W. West,
cond-mat/0012094.}

\def\<{\langle}
\def\>{\rangle}

\Title{cond-mat/0105027}
{\vbox{\centerline{Bilayer Quantum Hall System}
\smallskip
\centerline{As a Macroscopic Qubit}}}
\smallskip
\centerline{Takeshi Inagaki}
\smallskip
\centerline{\it Yamato Laboratory, IBM Japan}
\smallskip
\centerline{\it Shimotsuruma, Yamato-shi, Kanagawa, Japan}

\bigskip
\bigskip

\noindent
In the bilayer quantum Hall system, a spontaneously charge
imbalance state appears at the ground energy level.
Gap in the collective excitation energy makes it
stable against decoherence in macroscopic level.
This state behaves as a spin $1\over 2$ representation of $SU(2)$
and can be controlled by applying the interlayer voltage.
We suggest this system can be regarded as a macroscopic realization
of a qubit for a quantum computer.

\vskip 3cm
\noindent
\Date{May, 2001}

\newsec{ Introduction}
In recent years, a number of realization of a qubit for a quantum computer
have been proposed. Many of them use a microscopic object, like an electron
or a photon or a nucleus, as a carrier of quantum information.
For example, in the NMR liquid quantum computer 
\refs{\Cory,\GershA,\GershB}
which is the first realization of quantum computing on the Earth
\refs{\GershC,\Chuan,\JonesA,\JonesB}, an ensemble of nuclear spins is
manipulated. Though it has huge number of nuclei, quantum phase information
is preserved only within a molecule in which a set of nuclei is embedded
and computing is executed in every molecules. Then result is read as an 
average in a canonical ensemble of them. A kind of difficulty will exist
with this microscopic qubits when number of qubits becomes large.
It must manipulate an extremely large molecules with many nuclear spins.
But when the size of a molecule becomes large, it becomes sensitive to
external fluctuation and quantum phase information will be lost.
In addition, there is the limit in resolution of NMR and it is difficult
to read many spin degrees of freedom at once.

In this paper we investigate a possibility to realize a qubit with
a collection of electrons which are strongly correlated and preserve
phase information of collective modes. Now a day, many
phenomena are known in which the quantum effect is observed in
macroscopic level. An example is the superconducting circuit with
a Josephson junction (SQUID) in which  magnetic
flux caused by movement of electrons in a circuit is constrained by
a effective potential due to the Aharonov-Bohm effect. An application
of this system to quantum computer was discussed in
\refs{\mooij,\orlando,\ioffe}.
  
The quantum Hall effect is another novel example of
the macroscopic quantum effect \refs{\girvin}.
The root cause of appearance of quantum effect in macroscopic level is
excitation gaps in kinetic energy of state of electron.
At low temperature limit, higher excitations become negligible and states
in the lowest level are isolated from others. This suppresses the
kinetic degrees of individual electrons and modulation in electron density
(sound wave) and spin polarization (spin wave) are left as degree of
freedom of the system.
It is well known that there is no gapless density wave (phonon) excitation
in a system with translation symmetries.
On other hand, spin wave is a gapless excitation due to $SU(2)$ symmetry
breaking of a vacuum. In the case of a real spin, an energy gap to first
excitation is caused by the Zeeman interaction with external magnetic field
and a vacuum becomes stable.
At the ground energy level, we have two states with homogeneous electron
density and fully spin  polarized.
This spontaneous spin polarization is called quantum Hall ferromagnet
caused by the exchange energy of Coulomb interacting fermions.
If two layers of quantum Hall liquid with a filling factor $\nu={1\over2}$
are located near by each other, an electron
in one layer and  a hole in another layer couple and develop coherence.
In this system called bilayer quantum Hall system, a pair of electron
and hole behaves as a single particle with pseudo spin $1\over2$.
Here (pseudo) ferromagnet is observed
again and charges of two layers are imbalance spontaneously.
Pseudo spin wave excitation is gapless too and we can make a gap another
way than Zeeman term. By turn on a magnetic field parallel to a quantum
Hall surface, long wave excitations are suppressed and discrete
(quantized) excitation levels are remained.
We have two states fully pseudo spin polarized as the ground states and
due to gapes in excitation energy, they are stable against decoherence.
Degeneracy of these two are resolved by applying interlayer voltage,
and the bilayer quantum Hall system can be operated as a macroscopic
qubit of quantum computer.
In section 2 we quickly review the quantum physics of
the quantum Hall systems.
In section 3 we argues its application to a quantum computer.
Note that another application of the quantum Hall effect to
quantum computer was discussed in \refs{\privman}.

\newsec{ Quantum Hall Systems}
\subsec {An electron moving in a magnetic field}
Quantum Hall liquid is a system of electrons confined in a 2 dimensional 
surface and suffered under the strong constant magnetic field which is
orthogonal to the surface. This magnetic field forces movement of an
electron in a circle and its quantized excitation energy becomes gapped.
The Lagrangian for an electron is
$$
{\cal L}_{electron}={m \over 2}{\dot x}_i^2 + eB \epsilon_{ij} {\dot x}_i 
x_j,
$$
where $i,j=1,2$ are index for coordinates of 2 dimensional surface
in which electrons are confined. This problem is solved as
a 1 dimensional harmonic oscillator and its excitation energy known
as Landau level is
\eqn\landau {
{\epsilon}_n=(n+{1\over 2}){eB \over m}.
}
This gap in the energy levels causes a quantum phenomenon named quantum
Hall effect. In a typical experiment environment, its energy scale is
$\sim 100K$. Now we consider the limit in which the temperature is low
enough and the constant magnetic field $B$ is strong enough.
Where the energy gap becomes wide and each  Landau level can be treated
in separate. This makes it possible to ignore higher kinetic
excitations and explore physics in the lowest Landau level with $n=0$.
The Lagrangian becomes
\eqn\llllag {
{\cal L}_{LLL}=eB \epsilon_{ij} {\dot x}_i x_j.
}
Here, the conjugate momentum for $x_i$ is $p_i=eB \epsilon_{ij} x_j$ and 
quantization of this system make the space of coordinates noncommutative
$[x_i, x_j]={i\over {eB}}\epsilon_{ij}$.  Physical origin of this 
noncommutative nature is the Aharonov-Bohm effect on electrons
and truncation of the system to the LLL is compensated by this
noncommutativity. From this, an electron occupies
area $1\over eB$, and in other word, $eB$ of electrons can degenerate
in a unit area. A rate of this density and actual electron density
is called filling factor $\nu$ and characterizes a quantum Hall system.
To understand collective behavior of electrons
in the LLL, we move to the effective field theory description of
the system in next subsection.

\subsec {Collective excitation modes}
To move from the first quantized single electron picture to the second 
quantized field theory picture \refs{\susskind}, introduce fluid fields
$x_i(y)$ where $y_i$ are the 2 dimensional coordinates in which
the density of electrons becomes a constant ${\rho}_0$ over the surface.
Because of the Coulomb interaction between electrons, a solution
$x_i(y)=y_i$ is a stable vacuum solution and the quantum fluctuation is
described by $a_i$ as $x_i=y_i+a_i$. With this variable, density of
electrons is $\rho={\rho}_0 det |\delta_{ij}+\partial_i a_j|$.
We get the field theory Lagrangian from \llllag\ as
${\cal L}=eB{\rho}_0 \int dy^2 \epsilon_{ij} {\dot a}_i a_j$. This is
the 2+1 dimensional abelian Chern-Simons gauge theory \refs{\dunne}
in the temporal gauge. The gauge invariant form of the Lagrangian is
$$
{\cal L_{CS}}= {\nu} \int dy^2 \{
   \epsilon^{\mu\nu\rho} a_\mu \partial_\nu a_\rho
\}
$$
where $\nu=eB\rho_0$ is filling factor of quantum Hall system.
Naively, the Hamiltonian for this theory equals to zero and this theory
dose not describe any dynamics. This is expected because the kinetic
degrees of freedom of original electrons are removed from an effective
theory for the LLL. Still we may have finite degrees of freedom which
come from global structure of a device.
For accurate analysis, we need to take into account a boundary
condition of fields. If we consider a system on a compact Rieman surface,
there remain finite number of degrees of freedom depending on the topology
of the surface. Gauge symmetry of the theory is the area preserving
diffeomorphism and gauge transformation acts on $a$ as
$a_i \rightarrow a_i+\partial_i \phi$.
From the Cohomological argument, number of physical degree of freedom
is given by $dim H^1(\Sigma;R)=g$, where $\Sigma$ is the compact Rieman
surface on which the theory is defined and $g$ is number of genus on
it. Each of degrees has $\nu$ of degenerated states and vacuum degeneracy
of quantum Hall system on a compact manifold is $\nu^g$.
On a sphere $g=0$ there is no dynamics and the state stays on the 
ground. Physically, this means there can not exist any sound wave 
excitation
in the fluid of electrons propagating on a sphere or a disk, if the 
boundary
effect (edge excitation) can be ignored. Absence of phonon like excitation
in the LLL can be understand as the following. A wave of momentum $k$
requires movement of electrons spread over area $1\over{k^2}$. This area
contains $\nu eB\over{k^2}$ of electrons and each of them has kinetic 
energy
${k^2}\over{2m}$. This indicates whole energy of excitation is independent
of $k$ and more accurate calculation shows its value coincides with
Landau energy gap $\epsilon={\nu eB\over m}$. This means phonon excitation
pushes one electron to next Landau level.

We see, as far as only position of electrons are in consideration, there is
no gapless collective excitation in the fluid of electrons on a sphere.
But once spin of electron is taken into account, system develops collective
modes with macroscopic coherence.

\subsec {quantum Hall ferromagnet and bilayer quantum Hall system}
With spin degree of freedom of electron, 
another novel phenomenon, called quantum Hall ferromagnet
\refs{\sondhi, \girvin} ,
is observed where all spins are polarized in a direction at the ground
state. This is caused by the exchange energy of electrons. The electron
is a fermion and, if all spins are aligned in a same direction,
the wave function of multiple electrons must be anti-symmetric for
permutation of any two of electrons in the system. This requires
vanishing of the wave function when two electrons close to each other.
So the Coulomb interaction energy among electrons is lower than
other configuration where spins are not polarized.
To estimate excitation energy of the spin wave, it requires
knowledge of the LLL wave function.

Here, we shortly review the derivation of this excitation energy
according to \refs{\girvin}.
General form of the wave function for a single electron in the LLL is
\eqn\wavefunction {
\psi =f(z) e^{-{1\over 4}|z|^2}
}
where $z=\sqrt{eB}(x_1+ix_2)$ and $f(z)$ is a polynomial of $z$.
The innerproduct of two states in the LLL is given by
\eqn\innerprod {
\<{\psi}_1|{\psi_2}\>
=\int dz^2 f_1({\bar z})f_2(z) e^{-{1\over 2}|z|^2}.
}
To represent a plane wave with momentum $k$ in the LLL,
we should use an operator
$
e^{2i{\bar k}{\partial_z}}e^{ikz}
=e^{-{1\over4}|k|^2}e^{2i{\bar k}{\partial_z}+ikz}
\equiv \tau_k
$
instead of usual c-function $e^{ikz+i\bar k \bar z}$.
The spin wave field with a momentum $q$ can be written as
$$
S_{\alpha}(q)=\sum_i s_{\alpha,i} \tau_{q,i}
$$
with the spin raising operator $s_\alpha$ 
where $\alpha=1,2,3$ and $i$ indexes electrons in the system.
The contribution to the Hamiltonian by the Coulomb interaction
between electrons are written as
$$
H_{Coulomb}=\int dk^2 v(k) \rho_{-k} \rho_{k}
$$
where $v(k)=e^2\int dz^2 {1\over {|z|}} e^{ikz}$ is the Coulomb
interaction represented in the momentum space and the density operator
is $\rho_{k}=\sum_i \tau_{k,i}$.
Now we find that $S_{\alpha}(q)$ and $H_{Coulomb}$ are not commute.
This results raising of energy by the spin wave excitation of
momentum $q$ as
$$
[H, S_{\alpha}(q)]=\int dk^2 e^{-{1\over 2}|k|^2} v(k) sin^2 ({1\over 
2}({q \times k}))
\equiv \epsilon_q.
$$
This value becomes zero at $q\rightarrow 0$ that means
this is gapless excitation. Small energy fluctuation brings
the system unexpected states and this is very undesired
for quantum computing. In the case of a real spin with magnetic
moment, Zeeman interaction with external magnetic field
perpendicular to quantum Hall surface causes additional
gap in excitation energy and ground states becomes stable.

Aside from this spin wave excitation, there exists spin texture mode
called Skyrmion whose excitation energy equals
${1\over4} \epsilon_{\infty}$.
So any spin fluctuation costs finite energy and,
at low temperature, state of this system is bound to one of two fully
polarized ground states. This is the quantum Hall ferromagnet.

Because a Skyrmion excitation carries electric charge $\nu e$
and it is the lowest cost way to inject a charge into
quantum Hall system, Skyrmions appears in quantum Hall system
in which filling factor $\nu$ is slightly differed from 1 and
spin polarization is rapidly broken at $\nu\ne1$. 

Same phenomenon as ferromagnet is observed in the
bilayer quantum Hall system \refs{\wen,\moon,\ezawa,\girvin} where
two layers of quantum Hall liquid are located in parallel and
very closed to each other to turn on interlayer tunneling.
An electron in one layer and a hole in another layer are in
interlayer coherence and behaves as a particle with the pseudo spin.
Charge and a filling factor in each layer is not definite because
of tunneling and the difference of charges between two layers
is a value of a spin in the $z$ direction.
The system with two of filling factor $\nu={1\over 2}$ quantum Hall
layers behaves as the quantum Hall liquid with filling factor $\nu=1$
with pseudo spin $1\over2$. States of this system are bound to two grounds 
again and vacuum transition takes place by applying the interlayer
voltage to $z$ direction which appears in a Hamiltonian as
the Zeeman term. Direct experimental evidence for this phenomenon
was reported in \refs{\spiA} .
Here we hope to use Zeeman interaction to control a vacuum transition
and we can not fix it to make a gap in pseudo spin wave excitation energy.
Fortunately, it is possible to make gaps in excitation energy by applying
parallel magnetic field to bilayer surfaces \refs{\spiB} .
A spin wave excitated state with momentum $q$ is described as,
$$
\phi_q(x)=\sum_i e^{i2\pi qx_i} |i\>
$$
where $|i\>$ denotes a states with $i$th pseudo spin is inverted.
Now move this inverted pseudo spin for a distance $l$.
This operation causes movement of an electron in the lower
layer and movement of a hole in the upper layer.
This can be regarded as circle movement of
an electron across both layer. If parallel magnetic field is
applied, a quantum state picks phase factor
$e^{ieB_{||}dl}$ during this operation, where $d$ is separation
of two layers and $B_{||}$ is a parallel magnetic field.
Consistency requires
$$
\phi_q(x)=\phi_q(x+{{2\pi}\over{eB_{||}d}}).
$$
This constrains $q=eB_{||}dm$ where $m=1,2,3,..$ and wave excitation
lower than $eB_{||}d$ is inhibited.

\newsec{Macroscopic Qubit}
\subsec{Vacuum transition}
Now consider a system of bilayer Quantum Hall liquids with filling
factor $\nu=1$ on a sphere at low temperature limit.
In this theory, all kinetic excitations are frozen away and we have only
two ground states in which all of electrons are in one of an
upper or a lower layer. To calculate tunneling amplitude from one vacuums
to  another, semi-classical approximation can be used as
$$
\Gamma \approx e^{-S_{cl}}
$$
where $S_{cl}$ is the Euclidean Action at a bounce solution which is a path
connecting two vacuums with minimizing the Action.
We can construct a bounce solution from a Skyrmion solution with winding 
number equals 1. Skyrmion is a topological solution.
Let's remind that our theory is defined
on a $S^2$ and the spin filed is a mapping $S^2 \rightarrow SU(2)$.
This mapping are classified by counting how many times $S^2$ covers
$SU(2)$.
If this number equals to 1, a spin is up on a pole and down on a pole at 
oppose side
and at intermediate region near $r=\lambda$, a spin is rotated 90 degrees.
Explicit form of this solution in the polar coordinates $(r, \theta)$ is
$$
\eqalign {
s_x&={{2\lambda r cos(\theta)} \over {\lambda^2+r^2}} \cr
s_y&={{2\lambda r sin(\theta)} \over {\lambda^2+r^2}} \cr
s_z&={{r^2-\lambda^2} \over \lambda^2-r^2}.
}
$$
As mentioned before, its excitation energy is
${1\over 4}\epsilon_{\infty}$.

Now the bounce solution is start $\lambda=0$ and go to $\lambda=\infty$. 
During this process energy dose not depend on the value of $\lambda$
and $S_{cl}$ is proportional to $\delta t$ which is time required for
process. Thus transition amplitude from spin up to down state
within time $T$ is
\eqn\bounce {
\Gamma(T)={{\int_0^T dt e^{-{1 \over 4}\epsilon_{\infty} t}}\over
{\int_0^{\infty} dt e^{-{1 \over 4}\epsilon_{\infty} t}}}
=1-e^{-{1\over4}{\epsilon_{\infty}T}}.
}

\subsec{Qubit operation}
Let denote the global spin alignment operator $M_z$ whose eignestates
are $M_z|+\>=|+\>$ and $M_z|-\>=-|-\>$ where $|+\>$ denotes all spins up
and $|-\>$ denotes all spins down.
$M_z$ is a good observable in our theory. It is measured by
difference of electric charge between two layers and controlled by
electric field $\cal E$ caused by interlayer voltage
which couples to $M_z$ by the Zeeman term
${\cal H}=eN{\cal E} M_z$ where $N$ denotes number of electrons in
the system. Initial state can be prepared by applying $\cal E$ for
enough interval. Then change the sign of $\cal E$. After a certain
period which is determined from \bounce\ ,
we will get the state $|\Phi\>=cos(\theta)|+\>+ sin(\theta)|-\>$. 
But this is not full $SU(2)$ rotation and the state dose not represent
a universal qubit of \refs{\barenco}. 
To realize a qubit from this quantum mechanism, the method of \refs{\divi} 
can be used in which a qubit is constructed from three spins governed by
the Heisenberg interactions between every pair of them. To produce
Heisenberg interaction, it requires a tunneling term between two qubits
$M_{1,+}M_{2,-}+M_{1,-}M_{2,+}$ aside of $M_{1,z}M_{2,z}$. It will need
tactical allocation of devices and a subject of further study. After
accomplishing this, a product of three spin states rotates under $SU(2)$.
Now we obtain a qubit. In our system, a qubit is represented by
the difference of charges between two layers which consists of many
electrons and this affects other qubits via the electric field produced
by this charge imbalance.

\newsec{Acknowledgements}
The author thanks D. DiVincenzo in IBM T. J. Watson Research Center
for pointing out a mistake in the earlier version of this paper.

\listrefs
\end